\documentclass[10pt,letterpaper]{proc}

\usepackage[top=0.6in, bottom=1.0in, left=0.6in, right=0.6in]{geometry}

\RequirePackage{calc}
\usepackage{times}
\usepackage{graphicx}
\usepackage{parskip}  
\usepackage{epsfig}
\usepackage{wrapfig}
\usepackage{amsmath}
\usepackage{amssymb} 
\usepackage{algorithm}
\usepackage{algorithmic}
\usepackage{float}
\usepackage{url}

\newcommand{\bv}[1]{\boldsymbol{#1}}           

  \usepackage{cite}

\begin{document}
\title{Simulating the Dynamic Behavior of Shear Thickening Fluids}

\author{ \textbf{Oktar Ozgen \hspace{1cm} Marcelo Kallmann} \\
School of Engineering, University of California, Merced,
CA, 95343 \\
E-mail: \{oozgen,mkallmann\}@ucmerced.edu \\
\\
\textbf{ Eric Brown}\thanks{E. Brown contributed to the work while at UC Merced} \\
School of Engineering and Applied Science,
Yale University, New Haven, CT 06520 \\
E-mail: eric.brown@yale.edu 
}


\maketitle

\begin{abstract}
While significant research has been dedicated to the simulation of fluids,
not much attention has been given to exploring new interesting behavior that can be generated with the different types of non-Newtonian fluids with non-constant viscosity.
Going in this direction, this paper introduces a computational model for simulating the interesting phenomena observed in non-Newtonian shear thickening fluids, which are fluids where the viscosity increases with increased stress.
These fluids have unique and unconventional behavior, and they often appear in real world scenarios such as when sinking in quicksand or when experimenting with popular cornstarch and water mixtures. 
While interesting behavior of shear thickening fluids can be easily observed in the real world, the most interesting phenomena of these fluids have not been simulated before in computer graphics.
The fluid exhibits unique phase changes between solid and liquid states, great impact resistance in its solid state and strong hysteresis effects. 
Our proposed approach builds on existing non-Newtonian fluid models in computer graphics and introduces an efficient history-based stiffness term that is essential to produce  the most interesting shear thickening phenomena. The history-based stiffness is formulated through the use of fractional derivatives, leveraging the fractional calculus ability to depict both the viscoelastic behavior and the history effects of history-dependent systems.
Simulations produced by our method are compared against real experiments
and the results demonstrate that the proposed model successfully captures key phenomena  observed in shear thickening fluids.
\end{abstract}


\section{Introduction} \label{sec:introduction}

In the field of rheology, the mechanics of fluids is typically characterized by a viscosity function, where the viscosity is defined as the ratio of shear stress to shear rate in a steady state flow. For a Newtonian fluid like water, this viscosity is a single-valued constant, but in non-Newtonian fluids this viscosity may be a function of shear rate and other parameters.  Dense suspensions of hard particles are known to be shear thickening, meaning that the viscosity function increases with shear rate over some range.

\begin{figure*}[htbp]
  \centering
  \includegraphics[width=1\linewidth]{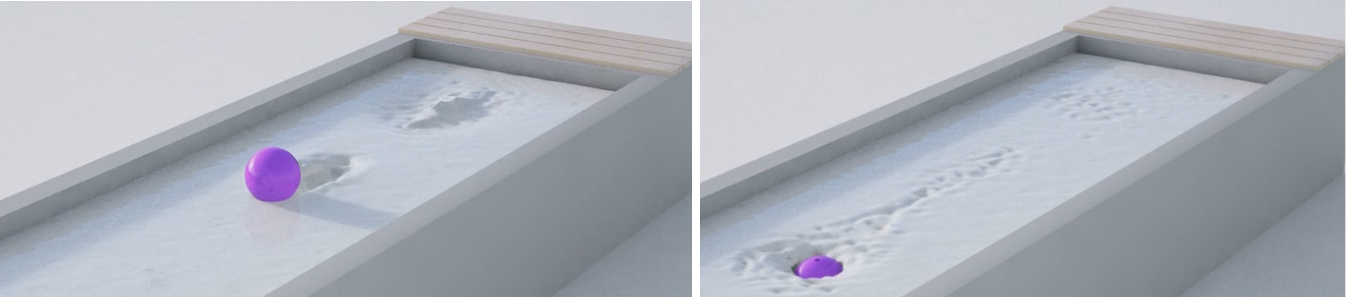}
  \caption{\label{fig:firstpic}
  This scenario illustrates the typical behavior of a shear thickening fluid: while the bowling ball  hits the fluid with high energy the bowling ball bounces (left image), and only when the energy starts to dissipate the bowling ball will then gradually sink, as in a standard fluid (right image).
           }
\end{figure*}

Shear thickening fluids (STFs) form a sub-category of non-Newtonian fluids exhibiting unconventional behavior that is not seen in other Newtonian or non-Newtonian fluids. These fluids harden when agitated by strong forces and soften in the absence of any forces. Moreover, they are history-dependent fluids in the sense that they keep their solid-like state for a given amount of time even after the forces are applied, i.e., their memory only fades away after some time. See Figure~\ref{fig:firstpic} for an example obtained with our simulation model.

The most known example of shear thickening fluids is probably the cornstarch and water mixture. The odd behavior of this fluid has recently attracted much public attention and videos showing experiments involving mixtures of cornstarch and water are widely spread over the Internet. In these experiments, it is possible to witness the dramatic impact resistance of the fluid to the point that a person can run on top of its surface without sinking~\cite{TimeWarp}. 

Another STF that has received some attention in several movies is quicksand. Innovative applications of STFs have also been developed. For example, the United States Army Research Laboratory has developed a ``Liquid Body Armor'', an armor suit that has layers of shear thickening fluid and Kevlar mixture. The fabric made of STF is highly resistant to penetration when impacted by a spike, knife or bullet, without compromising its weight, comfort or flexibility in normal conditions \cite{GizmagUrl} \cite{HowstuffworksUrl}. 

Despite the growing interest in the properties of shear thickening fluids, no computational models are known to date to simulate their complete behavior. 
For instance, while several models exist to simulate viscoelastic fluids, viscoelastic fluids and shear thickening fluids are non-overlapping subcategories of non-Newtonian fluids. 
Our proposed model in particular addresses a subcategory more specifically referred to as \emph{discontinuous shear thickening} in the literature,
which includes the most interesting phenomena observed in cornstarch and water mixtures (shown in Figures~\ref{fig:firstpic} and \ref{fig:monsters-compare}.)

Our contribution in this paper is significant to computer graphics in several aspects: 
it demonstrates new simulation effects that can be achieved with fractional derivatives, 
it shows that phenomena not previously simulated can be addressed with reasonably simple SPH-based techniques, and it introduces to computer graphics the unconventional behavior of an interesting class of materials.
The model we introduce in this paper is the first to produce visually successful simulations of the most interesting phenomena observed in shear thickening materials.

\section{Related Work}\label{sec:relwork3}

The simulation of shear thickening fluids involves concepts from different disciplines. We first review previous work in physics that discuss the behavior of shear thickening fluids. We then review a group of rheology and physics papers that use fractional calculus to describe viscoelasticity and the behavior of non-Newtonian fluids. Finally, we review previous work in computer graphics that are related to simulation of non-Newtonian or viscoelastic fluids.

\subsection{Physics and Rhelology}

While several models exist for shear thickening, none have been able to quantitatively predict the dramatic impact resistance observed in dense suspensions such as cornstarch and water.  One long-standing model for shear thickening attributes the increased dissipation rate of shear thickening to lubrication forces in the small gaps between densely packed particles \cite{Brady85} \cite{Wagner09}.  As the particles are sheared together the lubrication forces increase and the particles form `hydroclusters' that separate slowly because of the large lubrication forces.  

Another popular model connects shear thickening to dilatancy of a densely packed granular material.  When a densely packed suspension is sheared, the particles form force chains between particles in contact and the packing dilates (expands), which causes it to push against the boundaries of the system.  The boundaries can then apply a force that propagates back along the force chains.  In this model, the energy dissipation in the dilated state comes from solid friction between particles, with forces determined by the stiffness of the boundary.  This model has quantitatively explained the orders-of-magnitude increase in viscosity observed in shear thickening suspensions in shear rheometry measurements, where the relevant boundary is usually the suspension-air interface and the stress is due to surface tension \cite{Brown12}.

A model that is specific to impact response was developed by Waitukaitis and Jaeger \cite{Waitukaitis12}, where they suggest that colliding particles collect to form a transient solid front that develops underneath the impact site and propagates downward.  They found that for fast impacts in a deep pool of dense suspension, the deceleration of an impacting object could be accounted for by conservation of momentum as this solid region acts as an effective mass accumulating on the end of the impacter.


All of the aforementioned models predict forces that are at least an order-of-magnitude 
too small to explain all the observations of people running on top of dense suspensions,
and a physical mechanism that can account for the most dramatic impact resistance is still unknown. No physics-based simulations have yet been able to show such
dramatic impact resistance.

The importance of hysteresis (history effects) is recently emphasized for the correct description of the fluid. Deegan \cite{Deegan10} showed that indentations in the surface of a vibrated shear thickening fluid only become stable if there is hysteresis in the viscosity function.  In a different study, Kann et al. \cite{Kann11} suggested that oscillations in the velocity an object sinking in shear thickening fluid may also be due to hysteresis in the viscosity function.

\subsection{Fractional Calculus and Viscoelasticity} \label{sec:fracdefvisco}

Even though our model is the first to use fractional calculus for the simulation of shear thickening fluids in the realm of computer graphics, the approach comes from existing mathematical models developed in the physics and rheology fields. Fractional calculus is popular for describing the complex dynamics of relaxation, oscillation and viscoelastic behavior seen in non-Newtonian fluids. In general, fractional calculus is employed by replacing the time-derivatives of strain and stress in well-established ordinary models, by derivatives of fractional order. This results in a more appropriate model for fluids that exhibit both viscous and elastic behavior. 

A detailed literature review on the attempts of using fractional derivative models to non-Newtonian fluids in order to obtain analytic solutions is presented in \cite{Qi09}. The first work to adopt fractional derivatives to the problem of viscoelasticity was presented by Germant \cite{Germant38}. Following its introduction, Bagley and Torvik \cite{Bagley83} demonstrated predictive constitutive relations based on fractional derivatives for the viscoelasticity of coiling polymers. Makris and Constantinou \cite{Makris93} were able to successfully fit a fractional derivative based on the Maxwell model to the experimental data, hence showing that a fractional Maxwell model can replace the Maxwell model for silicon fluids. Heibiga \cite{Heibiga08} also used a one-dimensional fractional derivative based on the Maxwell model for the linear viscoelastic response of some polymers in the glass transition. Palade \cite{Palade99} reduced the constitutive equation for the incompressible fluid to the linear fractional derivative Maxwell model in the context of small deformations. 

\subsection{Computer Graphics}

Fractional derivatives have been introduced before in computer graphics with a deformation model for underwater cloth simulation. The approach achieves a history-based surface model that does not require the simulation of the body of water surrounding the cloth model~\cite{ozgen10tog}. In this paper, we employ the history effects of fractional derivatives for the purpose of simulating the interesting behavior of discontinuous shear thickening fluids.

Basic shear thickening behavior has been simulated before in previous work with a modified Herschel-Bulkley model~\cite{Yue15}. 
The Herschel-Bulkley model is an abstract  general form for a bulk rheology  that can include  shear thinning and shear thickening  in  a mathematical sense; however, it is well-known that the shear thickening that is described by 
this type of model is not able to simulate the behavior of discontinuous shear thickening fluids, such as the interesting phenomena exhibited by cornstarch and water~\cite{PhysRevE.81.036319,PhysRevE.87.042301}. 
The key physics that has to be taken into account is hysteresis in the constitutive relation, what is achieved by our model with the use of fractional derivatives.


Other relevant papers to our work are the ones addressing melting and flow of materials. Terzopoulos and Fleischer \cite{Terzopoulos89} developed a method that can simulate non-rigid objects that are capable of heat conduction, thermoelasticity and melting. The particles that constitute the material are connected by elastic springs and, in order to model the melting effect, the stiffness of the springs decreases as the temperature rises, and the particles break free when the springs disappear. 
Desbrun and Gascuel \cite{Desbrun95} also focused on the flow of soft substances. They employed a hybrid model by combining an implicit surface model with a particle system. Successful results were presented on the animation of soft materials that can undergo fusion.
Desbrun and Cani \cite{Desbrun96} demonstrated the first application of SPH in Computer Graphics for the animation of inelastic deformable bodies with a wide range of stiffness and viscosity.

Carlson and Turk \cite{Carlson02} took on the challenge of animating materials with both fluid and solid-like behavior. The target was on materials that can melt, flow and solidify. 
A modification of the marker-and-cell method was used to deal with the computational barrier of simulating highly viscous fluids, since the solidity of the materials was in fact achieved by simulating a highly viscous fluid. No elastic elements were added to the fluid equations.
 Goktekin et al. \cite{Goktekin04} presented a method that animates viscoelastic fluids such as mucus, liquid soap, toothpaste and clay. The method introduced elements of elasticity into an Eulerian implementation of the Navier-Stokes equations. The total strain is defined as the sum of elastic and plastic strains. 

Clavet et al. \cite{Clavet05} developed a powerful and simple method that can efficiently and realistically simulate viscoelastic fluids. The method combines a simplified SPH with a mass-spring system. The SPH is  responsible for the fluid behavior while the springs are responsible for the elasticity and plasticity effects that make the material more solid-like.  The method computes an additional pressure term named near-pressure, along with the conventional pressure term calculated through SPH. This term is used to solve the clustering problem which is a common problem in SPH implementations. Furthermore, the nearby particles are also connected to each other by springs. These springs have varying rest lengths and  are removed when the distance between a pair of particles is greater than a threshold. The use of adaptive springs along with the double density relaxation method results in a very strong animation tool for viscoelastic models. This method constitutes the starting point for our proposed shear thickening fluid animation model.

In conclusion, shear thickening fluids have no known complete physical model and their simulation is an interesting animation problem. At the same time, the ability of fractional calculus to describe history-based behavior is well established in rheology. Motivated by these facts, this paper builds on the non-Newtonian fluid models in computer graphics and introduces a fractional calculus based model to efficiently simulate the interesting behavior of discontinuous shear thickening fluids.

\section{Proposed Model}\label{sec:model3}

The starting point of our shear thickening simulation model is based on the model presented by Clavet et al. \cite{Clavet05}. We give below a summary of this model, and we then focus on describing our solution for achieving the shear thickening behavior.

\subsection{Preliminaries}

Clavet et al. \cite{Clavet05} simulate viscoelastic fluids using a Lagrangian model that combines SPH forces with a dynamic mass-spring system. The density of a particle is computed based on the distance to other particles in its vicinity. The pair-wise distances are smoothed through a smoothing kernel function. The densities are then used to calculate the pressures at particle locations. Two density terms namely near-density and far-density are calculated in slightly different ways and this procedure is called double density relaxation. The additional near-density term is employed to counteract the clustering problem and to create better surface tension. The viscosity force between two particles is calculated based on the difference between their velocities, in accordance with the Navier-Stokes equations. The difference in velocities has both linear and quadratic effects. The elasticity component originates from the spring-mass system. 

Particle pairs that are close to each other up to a given distance are connected to each other by elastic springs, and the spring force is given by: 

\begin{equation} \label{eq:hookespring}
\begin{array}{c}

\bv{F}_s = - k_{min} \: \bv{x},

\end{array}
\end{equation}

where $k_{min}$ is the spring constant. The springs only exist between pairs of particles that are closer to each other than a given distance. If two particles move farther away from each other, then the spring connecting them is removed. The rest lengths of the springs are also updated at every simulation step. The addition and removal of springs along with the rest length updates create the effect of plasticity. The displacements originating from the SPH and spring forces are applied through a prediction-relaxation approach. For more details about this model we refer the reader to the work of Clavet et al.~\cite{Clavet05}.


\subsection{Shear Thickening Behavior}

The existence of viscosity, elasticity and plasticity terms lead to the simulation of viscoelastic materials or non-Newtonian fluids. However, these components are not sufficient for creating a shear thickening fluid. 

Shear thickening fluids have two very important additional characteristics when compared to viscoelastic fluids. First, they change their state from liquid-like to solid-like when subjected to great forces. As long as forces are present, the solid-like state is preserved. In the absence of forces, the material will soften and return to its liquid-like structure. Second, shear thickening fluids are history-dependent materials.  The present state of these materials is defined in terms of past states, and the materials remember the past forces that affected them up to some extent. When a shear thickening fluid is in its solid-like state and all forces are ceased, it will take some time to go back to its liquid-like state. Likewise, when the fluid is in its liquid state and is agitated by mild forces, it will slowly move from liquid to solid state over time.

In our proposed approach, the most critical factor that will transform a viscoelastic fluid into a shear thickening fluid is the inclusion of history-based spring elements. These elements have a history-based stiffness constant that injects the information of the past into the system. The history-based stiffness terms are achieved by multiplying the stiffness constant of the dynamic springs by the magnitude of the fractional derivative of the position of the connected particle. The history-based spring element is given by:


\begin{equation} \label{eq:neweq0}
\begin{array}{c}

\bv{F}_{hist} = - k_{hist} \: \|D^q\bv{x}\| \: \bv{x},

\end{array}
\end{equation}

where $k_{hist} \: \|D^q\bv{x}\| $ is the history-based stiffness, $D^q\bv{x}$ is the $q^{th}$ order derivative of the position and $q$ is a non-integer between 0 and 1. 

The approach used to achieve the history effects relies on the properties of fractional calculus. As discussed in Section~\ref{sec:fracdefvisco}, fractional calculus is widely used for the description of viscoelastic phenomena and it provides an appropriate tool for the simulation of history effects. 

The reason for achieving history effects can be clearly seen by the fractional derivative formulation, which in its computational form involves a summation of past integer derivative terms. 
The numerical formulation adopted in this paper is based on the Riemann--Liouville integral (see Appendix).
A second order numerical solution of the fractional derivative of order $q$ for values $0<q<1$ is given by Soon et al.~\cite{Coimbra05}, and it reads:


\begin{equation}\label{eq:fd2005eq}
D^{q}\bv{x}_n  = \frac{\Delta t ^ {1-q}}{\Gamma(3-q)} \sum_{p=0}^n {a}_{p,n} D^{1}\bv{x}_p,
\end{equation}



\begin{equation*} \label{eq:weightsstf}
\begin{split}
{a}_{p,n} =
\begin{cases}
&(n-1)^{2-q} - n^{1-q} (n+q-2)  \;\;\; \text{if } p = 0,\\
\\
&(n-p-1)^{2-q} - 2(n-p)^{2-q} + (n-p+1)^{2-q} \\
& \text{if } 0 < p < n,\\
\\
&1 \;\;\; \text{if } p = n,
\end{cases}
\end{split}
\end{equation*}

where $q$ is the derivative order such that $0 < q < 1$, $n$ is the index of the most recently computed timestep, ${a}_{p,n}$ is the weight of the 
past timestep $p$ computed at the current timestep $n$, $\Gamma$ is the Gamma function, and $D^{1}\bv{x}_p = \bv{v}_p$ 
is the velocity of the particle at past timestep $p$.

\begin{figure}[t]
  \centering
  \includegraphics[width=1\linewidth]{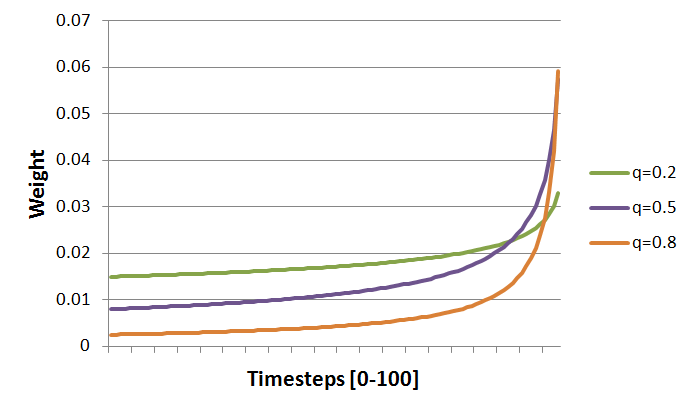}
  \caption{\label{fig:weights-q020508}
  Weights produced by fractional derivatives of order $0.2$, $0.5$ and $0.8$.
           }
\end{figure}

The fractional derivative of the position at the current time step is thus computed based on the velocities of the past, and the past velocities are prioritized by weights. The near past has more influence on the current state than the far past. The evolution of the weights generated by the fractional derivative operator is shown in Figure \ref{fig:weights-q020508} for $q$ values of $0.2$, $0.5$ and $0.8$.

We compute the total spring force as the sum of forces generated by the history-based spring element and the regular elastic spring:

\begin{equation} \label{eq:addit}
\begin{array}{c}

\bv{F} = - k_{min} \: \bv{x} - k_{hist} \: \|D^q\bv{x}\| \: \bv{x}.

\end{array}
\end{equation}

Substituting Equation \ref{eq:fd2005eq} into Equation \ref{eq:addit} and rearranging, we obtain:


\begin{equation} \label{eq:neweq1}
\begin{array}{c}

\bv{F} = -1 \: \left( k_{hist} 
 \:\dfrac{\Delta t ^{1-q}}{\Gamma(3-q)}\:
\left\| \sum\limits_{p=0}^{n} a_{p,n} \bv{v}_p \right\|
+ k_{min} \right) \:\bv{x},

\end{array}
\end{equation}

where $k_{hist}$ is the control parameter for the history-based spring, 
$a_{p,n}$ is the weight of the past timestep $p$ computed at the current timestep $n$, 
$\bv{v}_p$ is the velocity of the particle at the past timestep $p$, 
$k_{min}$ is the stiffness constant of the elastic spring, and 
$\bv{x}$ is the vector of elongation or compression. 

The above formulation successfully models both of the desired features of shear thickening fluids. 
First, the spring immediately becomes very stiff when the velocity $\bv{v}_p$ dramatically increases in the near past, what results in the mass-spring system exhibiting solid-like behavior in regions where great impact forces are applied.
Second, the spring will gain memory thanks to the weights $a_{p,n}$ that apply to the velocities of the past. 
This results in the overall stiffness of the springs gradually rising under continuous application of forces, and gradually decreasing under the absence of forces. The overall behavior achieved by the spring-mass system will lead to the shear thickening fluid slowly changing its phase both from liquid to solid and from solid to liquid. 

Although $k_{hist}$ and $k_{min}$ are both denoted with the letter $k$ in accordance with the Hooke's law, they serve different purposes. Constant $k_{min}$ is the stiffness of the elastic spring and it specifies the minimum stiffness that the spring can have. The history-based stiffness originating from the fractional derivative is an additional term on top of the default stiffness. In the absence of great velocities, notice that the first term in the parenthesis of Equation \ref{eq:neweq1} vanishes and the system behaves exactly like a regular viscoelastic fluid as described in \cite{Clavet05}. However, when the particle gains instantaneous or cumulative high velocity over time, the first term adds extra stiffness to the spring. Constant $k_{hist}$ is the parameter that controls the amount of extra stiffness that will be created due to the velocity increase. Constant $k_{hist}$ can be used to give to the animator the ability to implicitly control the maximum stiffness that the system can gain. In cases where the integration method cannot keep the system numerically stable because of excess of stiffness, $k_{hist}$ can be fine-tuned to keep the simulation stable. In our experiments, we have set $k_{hist}$ so that the added stiffness can increase up to approximately ten times $k_{min}$.


\begin{figure*}[htb]
  \centering
  \includegraphics[width=0.9\linewidth]{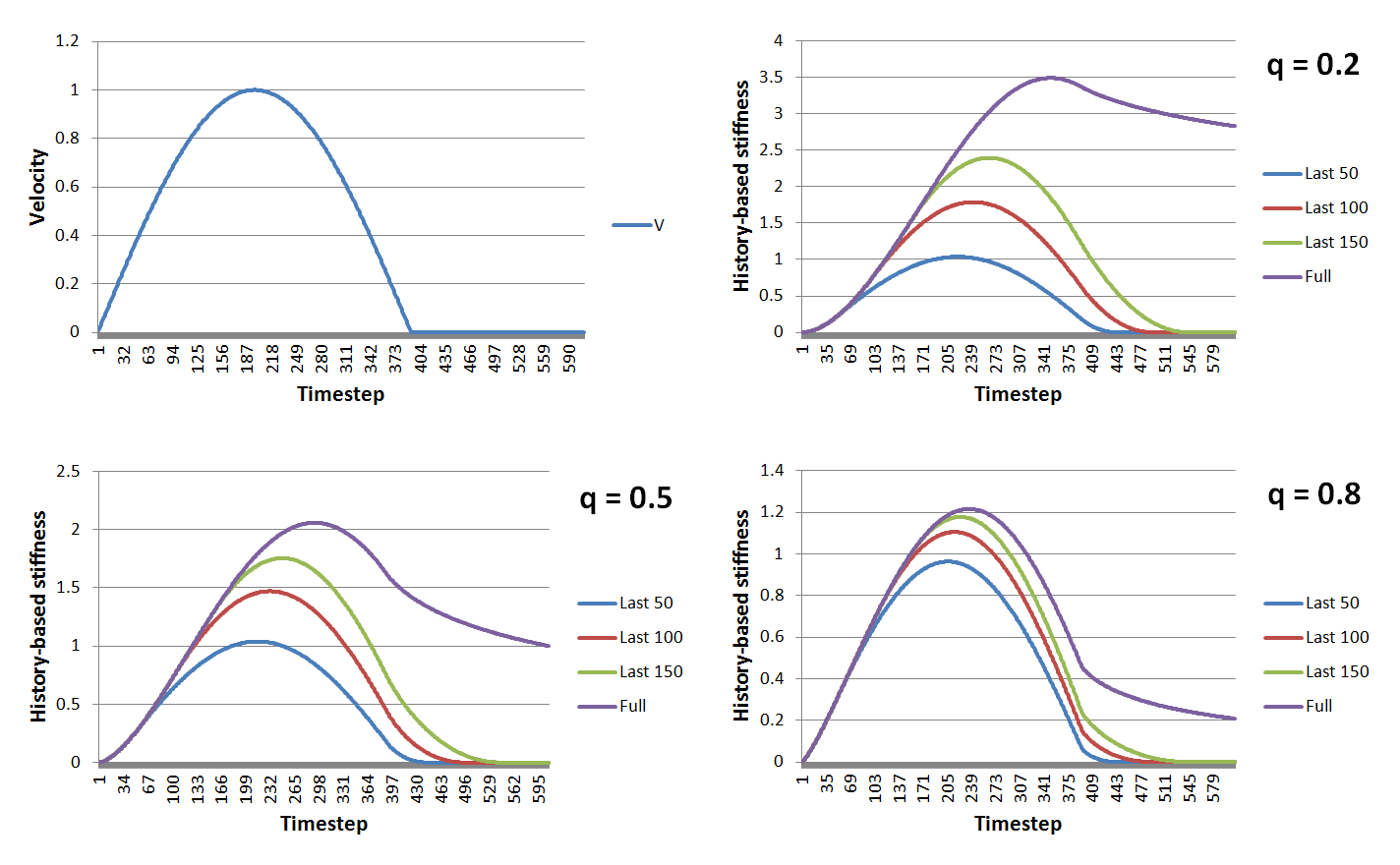}
  \caption{\label{fig:graph1}
  The upper-left graph shows the velocity  of a single particle subjected to a temporary velocity increase due to an impact. The next three graphs (in left-right, top-down order) show the evolution of the history-based stiffness values of the spring connected to the same particle, when its velocity changes according to the upper-left graph. Curves for various $d$ values are given in each graph. The second, third and forth graphs are based on the order of the fractional derivative  $q = 0.2$, $q = 0.5$ and  $q = 0.8$ respectively.
           }
\end{figure*}

\begin{figure*}[htb]
  \centering
  \includegraphics[width=0.9\linewidth]{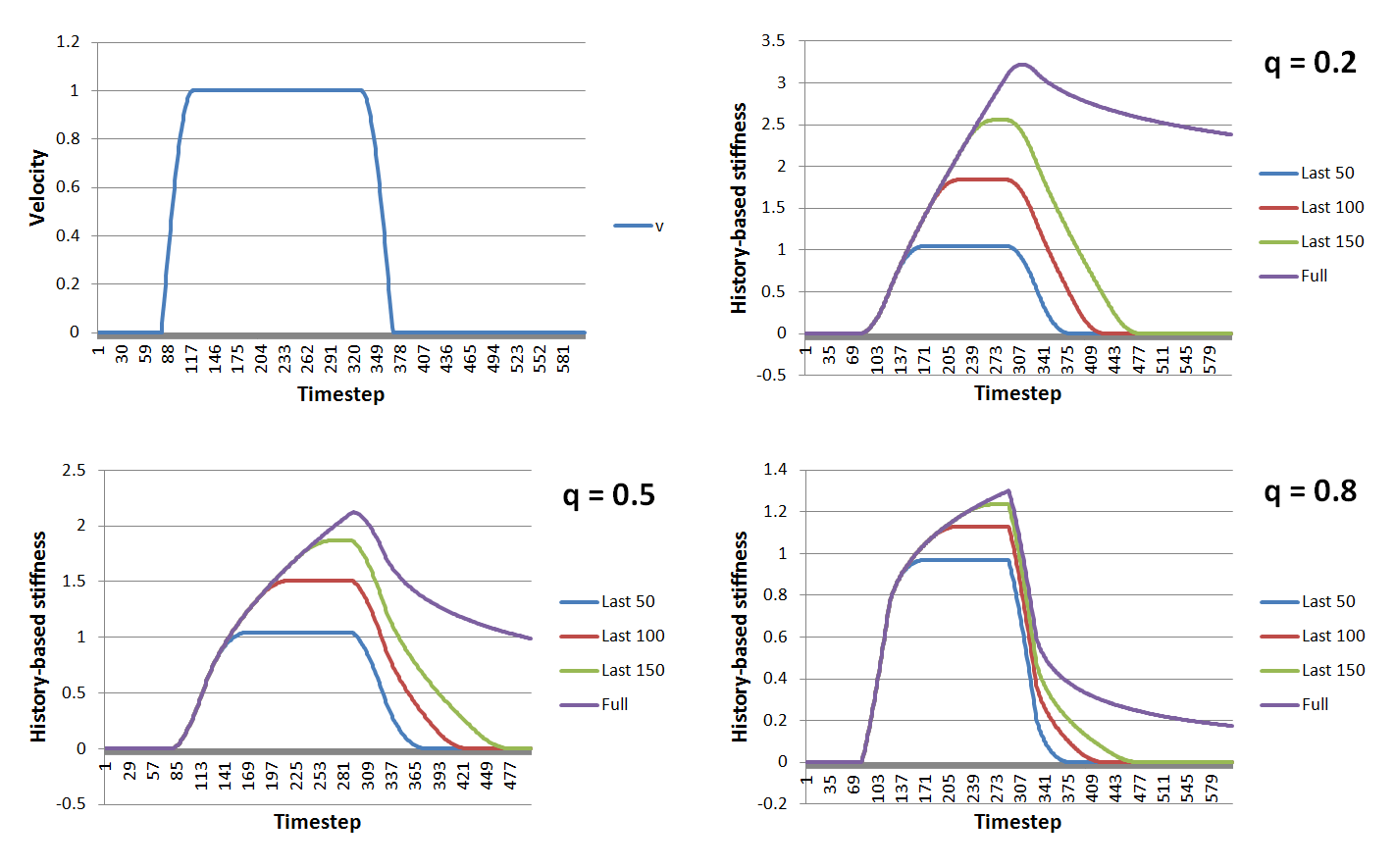}
  \caption{\label{fig:graph2}
    The shown graphs present the same plots as in Figure~\ref{fig:graph1}, except for the upper-left graph which shows the velocity  of a single particle moving with constant velocity for a given period of time.
}\end{figure*}

One important numerical drawback of the fractional derivative formulation is that it is a global derivative over the whole history of the particle. Clearly, this would pose performance problems as the simulation moves forward since the cost of calculating each simulation step would be always greater than the cost of the previous step. Therefore we limit the amount of history to be considered by changing the start index of the sum in Equation \ref{eq:neweq1} from $p=0$ to $p=n-d$, where $d$ is the number of time steps in the past that we would like to take into account. Parameter $d$ is also a useful parameter to model the characteristics of the material. If the material being modeled has long-term memory, it makes  sense to use a large number of past states. On the other hand, if the material has short-term memory, a small value for $d$ should be chosen.

We have investigated the behavior of a single particle under the influence of our history-based spring element in order to analyze the evolution of the history-based stiffness with respect to the particle's velocity magnitude. Figures \ref{fig:graph1} and \ref{fig:graph2} present several history-based stiffness values generated by various memory settings (various $d$ values) and for history-based spring terms with various $q$ values. Figure~\ref{fig:graph1} demonstrates how the history-based stiffness changes when the particle gains a temporarily high velocity due to an impact. Figure~\ref{fig:graph2} shows the evolution of the history-based stiffness when the particle moves with a steady velocity for a longer period time. It is possible to observe that the history-based stiffness increases more and lasts for a longer period when more states in the past are considered. Furthermore, it can be seen in the graph that limiting the number of past timesteps naturally creates a maximum limit for history-based stiffness, leading to obvious advantages in terms of controlling the numerical stability. In real experiments with shear thickening fluids composed of cornstarch and water we have observed that the material has the approximate memory of one second. Therefore we have chosen $d=1 / \Delta t$ in our simulations.

Another important modeling decision is the choice of the fractional derivative order. Equation 3 
is defined for values of $q$ such that $0<q<1$. The evolution of the weights generated by the fractional derivative for different values of $q$ is presented in Figure \ref{fig:weights-q020508}. It can be observed that as $q$ gets closer to $1$, the history effects tend to disappear. As the value of $q$ gets close to $0$, the effect of the past time steps tend to become similar. Both extreme cases do not seem to be useful for our problem. From previous work on fractional derivatives on memory systems we know that the half-derivative ($q=0.5$) is the best choice for describing a number of phenomena in rheology \cite{Coimbra98} \cite{Coimbra03} \cite{Coimbra04} \cite{Lesperance05} \cite{Coimbra05}. We have therefore used $q=0.5$ as the fractional derivative order in our experiments.
The value of $q=0.5$ was also experimentally observed to produce best results.

Different values of $q$ can be always specified by animators in order to experiment with different history effects. For the sake of examining the different behaviors that can be achieved, an analysis of the history effects generated with $q=0.2$, $q=0.5$, and $q=0.8$ is also presented in Figures \ref{fig:graph1} and \ref{fig:graph2}. 
We also present results 
demonstrating a longer-lasting solidification effect that can be achieved with $q=0.2$ (Figure~\ref{fig:bb2}).

Considering all discussed parameters and adapting Equation \ref{eq:neweq1} to the case of two particles connected by a spring, the total force between two given connected particles is:

%
%
%

\begin{equation}\label{eq:neweq2}
\begin{split}
\bv{F} = &-1 \: \left(k_{hist} 
 \:\dfrac{\Delta t ^{0.5}}{\Gamma(2.5)} \: 
\left\| \sum\limits_{p=n-d}^{n} a_{p,n} \:
(\bv{v}_{pi}-\bv{v}_{pj})
\right\|
 +k_{min} \right) \\
&\: \dfrac{\bv{x}_i-\bv{x}_j}{\|\bv{x}_i-\bv{x}_j\|}\: (\|\bv{x}_i-\bv{x}_j\| - L),
\end{split}
\end{equation}

where $\bv{v}_{pi}$ and $\bv{v}_{pj}$ are the velocities of the particles $i$ and $j$ at the past timestep $p$ and $L$ is the rest length of the spring. As part of the employed prediction-relaxation integration scheme \cite{Clavet05}, the displacement generated by the spring force is computed by multiplying the spring force by $\Delta t^2$, and then the displacement is applied to the positions of the particles.

\subsection{Boundary Conditions}

Experiments with real shear thickening fluids show that shear thickening can support objects with extreme masses. A bowling ball can jump and roll, and a person can literally walk on a shear thickening fluid while the fluid is in its solid-like state. Experiments in physics and rheology stress the importance of solid container walls in other shear thickening behavior~\cite{Brown12}. The force applied to the surface is eventually transmitted to the boundaries of the container and solid walls help the fluid to support even greater masses. 

Our network of history-based springs can simulate this behavior up to some extent. To improve the overall results an additional technique is employed for making sure that the particles at the boundary of the fluid are highly sticky to the boundary walls. For a particle at the boundary, its velocity is separated into its normal and tangential components. We completely cancel the normal velocity component that is perpendicular to the wall. We also lower the tangential velocity component in a proportional way to the average of the history-based stiffness of the springs that the particle is connected to. In this way, when the fluid exhibits solid-like behavior, the particles at the boundaries are almost fixed. As the stiffness of the whole system decreases, the particles at the boundaries become less sticky. 

\subsection{Summary of Overall Algorithm}

The main steps of the overall simulation algorithm 
are summarized in Algorithm \ref{alg:mainalgstf}. 

Time integration is computed through prediction-relaxation. In the prediction-relaxation approach, first the forces originating from gravity and viscosity are calculated, and then the velocity is updated based on these forces. The particle is then virtually moved to its predicted position based on the current velocity. At this predicted position, the density forces of SPH and the spring forces that account for the elasticity are calculated and the displacement due to these forces is immediately applied to the particle. 

Lastly, collisions between the particles and the walls of the container are resolved and the positions of the particles are corrected. The velocities of the particles are implicitly calculated at the end of the timestep.

\begin{algorithm}
\caption{Simulation Step}\label{alg:mainalgstf} 

\begin{algorithmic}[1]
\algsetup{linenodelimiter=.}
  \STATE // Apply gravity:
  \FORALL{particle $i$}
    \STATE $\bv{v}_i \leftarrow \bv{v}_i + \Delta t \bv{g}$
  \ENDFOR
	\STATE // Update velocities with viscosity forces:
	\STATE ApplyViscosity();
	
	\FORALL{particle $i$}
    \STATE $\bv{x}_i^{prev} \leftarrow \bv{x}_i$     // Save previous position
    \STATE $\bv{x}_i \leftarrow \bv{x}_i + \Delta t \bv{v}_i$ // Advance to predicted position
  \ENDFOR

	\STATE // Add or remove springs, or change rest lengths if necessary:
	\STATE AdjustSprings();
	\STATE // Apply displacements due to history-based spring elements:
	\STATE ApplySpringDisplacements();
	\STATE // Apply displacements due to pressure forces:
	\STATE DoubleDensityRelaxation();
	
	\STATE // Use previous positions to compute next velocities:
	\FORALL{particle $i$}
    \STATE $\bv{v}_i \leftarrow (\bv{x}_i - \bv{x}_i^{prev}) / \Delta t$
  \ENDFOR

	\STATE // Take into account the boundary conditions:
	\STATE DetectAndProcessBoundaryParticles();

\end{algorithmic}
\end{algorithm}

\begin{figure*}[ht]
  \centering
  \includegraphics[width=1\linewidth]{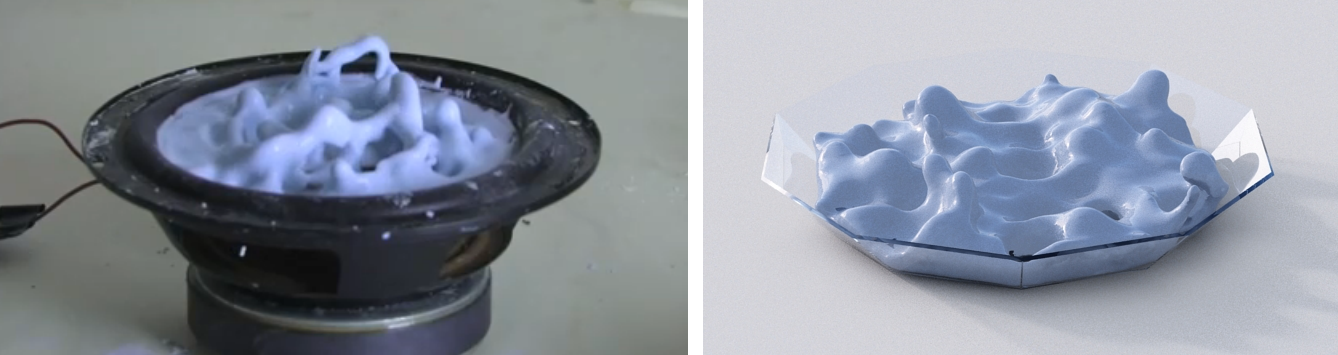}
  \caption{\label{fig:monsters-compare}
  Vibration experiment with a shear thickening fluid. The images show a comparison between the real experiment results \protect\cite{bendhoward} (left image)  and our results obtained in simulation (right).
           }
\end{figure*}

\begin{figure*}[ht]
  \centering
  \includegraphics[width=1\linewidth]{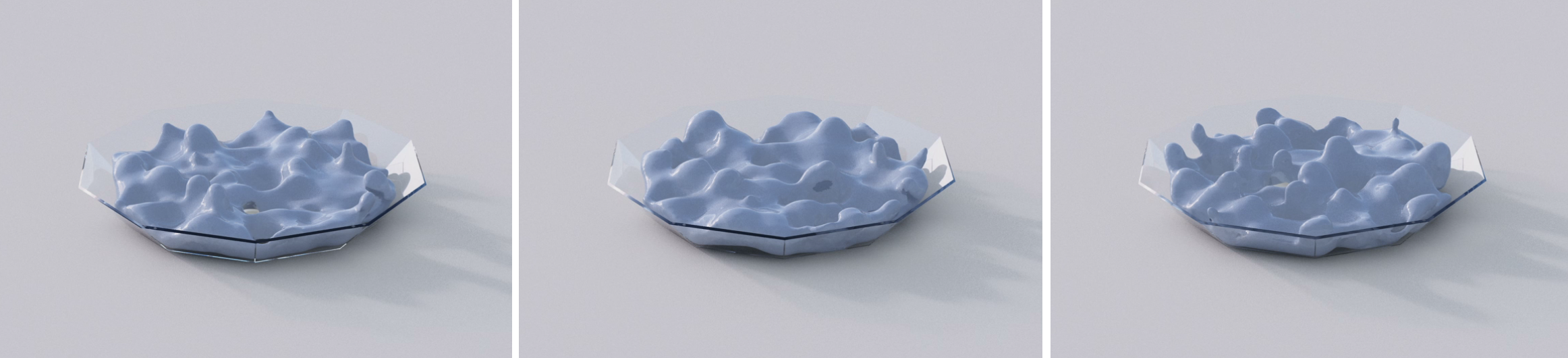}
  \caption{\label{fig:monsters-3snap}
  The snapshots from the vibration experiment with our shear thickening fluid model show the expected finger-like formations with various shapes. These formations do not appear if the history terms are removed.
           }
\end{figure*}

\section{Results and Discussion}\label{sec:results3}

There is an abundance of real experiments involving shear thickening fluids, namely with cornstarch and water mixtures, that show us how the material behaves in different conditions. We evaluate our model by visually comparing our simulations with these real experiments. 




In our first experiment we have simulated a scene where a bowling ball is rolled on the surface of the shear thickening fluid. The real version of this experiment is important in many aspects. First, it shows that the fluid can support a high mass object such as a bowling ball and that the solid-like phase of the material can even make the bowling ball slightly bounce on the surface. Second, the bowling ball can roll on the fluid surface as if rolling on a solid surface. Third, even when the motion of the bowling ball ends, it still stays on the surface of the fluid for a given time, before it slowly starts to sink. 

\begin{figure*}[htb]
  \centering
  \includegraphics[width=0.246\linewidth]{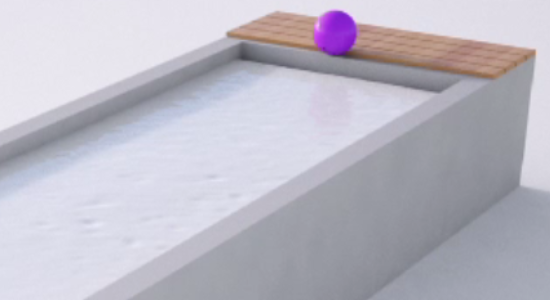}
  \includegraphics[width=0.246\linewidth]{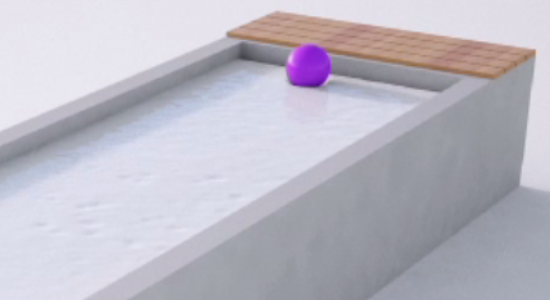}
  \includegraphics[width=0.246\linewidth]{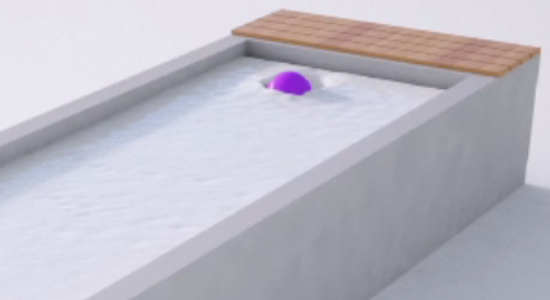}
  \includegraphics[width=0.246\linewidth]{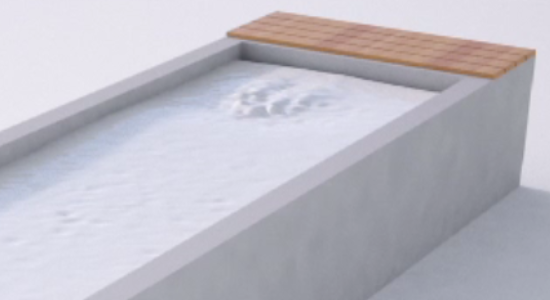}
  \caption{\label{fig:nohist}
    Without the history terms the bowling ball quickly sinks as in a normal fluid.
    }
\end{figure*}

\begin{figure*}[thb]
  \centering
  \includegraphics[width=0.62\linewidth]{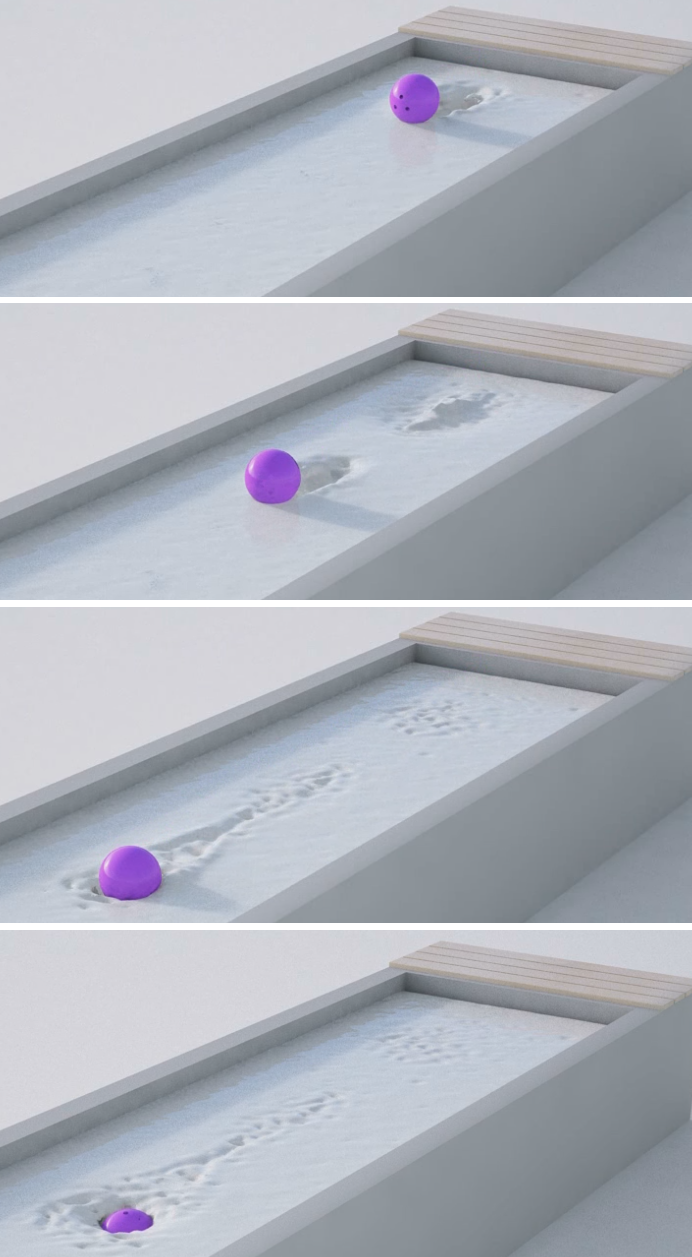}
  \caption{\label{fig:bb}
    The first image (in top-down order) shows the first moment of impact where the shear thickening fluid successfully carries the mass of the bowling ball and makes it bounce. The second image shows the second moment of impact after the bounce. The bowling ball still won't sink but slows down. The third image shows the moment where the bowling ball comes to a stop. Since the fluid is still in its solid-like state, the bowling ball will not immediately sink. The fourth image shows the end of the simulation where the history-based stiffness starts to vanish and the fluid transitions to fluid-like state, finally resulting in the sinking of the bowling ball.}
\end{figure*}

The bowling ball scene clearly shows the history effects in action in the mechanics of the fluid, and our model is able to successfully replicate in simulation the same behavior observed in real experiments. The fluid immediately solidifies under the high impact of the ball, acting somewhat like a spring mattress. It then keeps its solid-state for a while, letting the bowling ball roll on the surface, and when the ball stops the fluid gradually softens and enters into its liquid state allowing the ball to sink. These phases are illustrated by the snapshots presented in Figure~\ref{fig:bb}.
Without the history terms the ball just sinks, as shown in Figure~\ref{fig:nohist}.

\begin{figure*}[htb]
  \centering
  \includegraphics[width=0.99\linewidth]{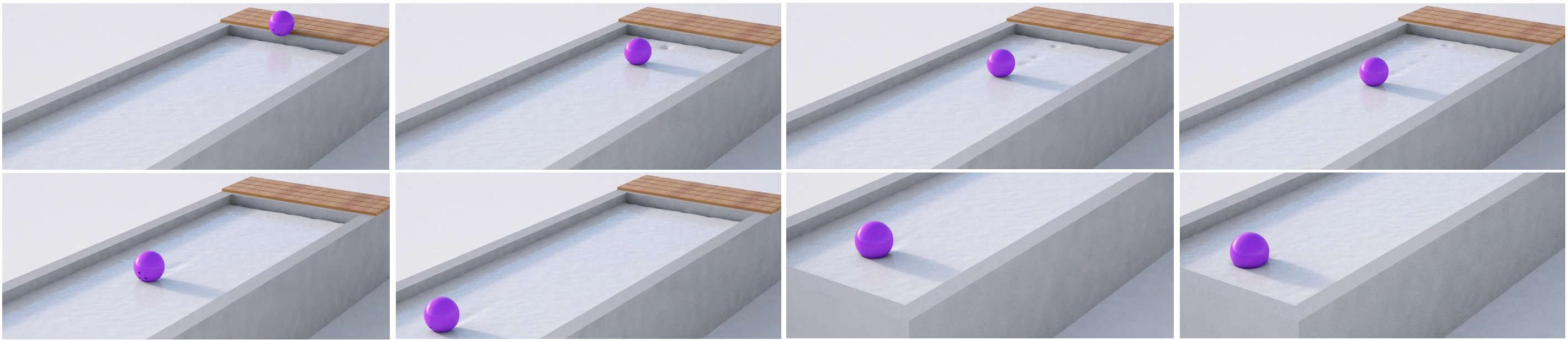}
  \caption{\label{fig:bb2}
    These snapshots show the long-lasting solidification effect achieved using a full history (i.e. using all the past timesteps) along with $q = 0.2$.
    }
\end{figure*}

In our second experiment we have simulated the shear thickening fluid subjected to vibration forces. In the real experiments the fluid is placed on a speaker creating vibrations that are transmitted to the fluid. After some time the fluid slowly starts changing its phase from liquid-like to solid-like. First the fluid forms holes and starts clustering at some random regions. Then the clustered regions slowly grow and form finger-like formations. The finger-like formations gradually get taller until they either merge with other parts of the fluid or break off from the main fluid body. 
Figure~\ref{fig:monsters-compare}-left shows a snapshot of the real experiment.

We have performed a similar experiment in simulation with a fluid with 10K particles. We have placed equally distributed vibration sources at the bottom of the container and we have added random changes in the vibration frequency throughout the simulation. 
Figure~\ref{fig:monsters-compare} compares the behavior observed in the real experiment against the results obtained in simulation with our method.
Additional results are presented in Figure~\ref{fig:monsters-3snap}, showing that our fluid model with history effects is able to successfully simulate the expected finger-like formations.

We have also compared the fluid with history-based springs against the fluid with no history terms. While, the vibration forces cause an accumulated history-based stiffness on the springs and the particles naturally start to climb on each other, the fluid without history terms could not achieve any of the finger-like formations. 

Our vibration experiment shows that finger-like formations require history terms, in agreement with the model of \cite{Deegan10}. The obtained results are the first to confirm that prediction in a dynamic simulation.  The fact that several of the phenomena associated with shear thickening have been simulated for the first time with a model with strong history effects suggests that hysteresis may be more generally important to shear thickening than previously indicated. These effects have been ignored in the standard models for shear thickening such as hydroclustering~\cite{Brady85} or dilatancy~\cite{Brown12} and this insight may lead to improvements of those models so that they can finally predict some of the interesting phenomena associated with shear thickening.

We have also tested the effects of increasing the number of past timesteps used to compute the fractional derivative (parameter $d$ in Equation \ref{eq:neweq2}) in conjunction with the history term $q$. In order to achieve a long lasting solidification, we used a full history (i.e. using all the past timesteps) along with $q = 0.2$. The expected behavior from this combination of parameters is shown in the upper right graph of Figure \ref{fig:graph1}. The simulation of this scenario matched our expectations and lead to a stiffer fluid after impact. The fluid took relatively more time to go back to its liquid state and a long lasting solid state is achieved. See  Figure~\ref{fig:bb2}.

We have also analyzed the parameter space of our model. The main significant parameter is the fractional order $q$. In our model when $q = 0.5$ (half-derivative) the history effects are most visible. The other variations of the parameter $q$ showed different in-between behaviors between solid and liquid. 
These variations are numerically demonstrated and the behavior acquired from different values of $q$ are compared in Figures \ref{fig:graph1} and \ref{fig:graph2}. 

The second important parameter, $d$, is the number of timesteps used to compute the half derivative. This parameter plays an important role in controlling the duration of the transition from solid to liquid state. We kept this parameter fixed in the first two animations in order to demonstrate the effects of the fractional order $q$. However, in our last animation (Figure~\ref{fig:bb2}), we  show that a long lasting solid state can be achieved by using a long history (i.e. very high $d$ values). The presented examples cover the most visually interesting combination of parameters, and a larger number of possibilities are presented numerically in the graphs of Figures \ref{fig:graph1} and  \ref{fig:graph2}.

In terms of performance, the running time obtained with our system is approximately 1.5 seconds per frame for a simulation of 20K particles in an Intel Core(TM) i7-2600K 3.4 GHz computer. The added computation time by the inclusion of history-based spring forces is minimal and these terms do not affect the complexity of the algorithm. The weights of the fractional derivative term can be precomputed and used in combination with the stored past velocities. Equation \ref{eq:neweq2} requires the algorithm to store the past velocities of the particles only up to a given number of timesteps, leading to an extra memory space requirement that is linear to the considered history size.
For a history of $d$ timesteps, $d$ velocity vectors (and pre-computed scalar weights) are stored.
Table~\ref{tab:perf} summarizes computational times obtained with our system.

\begin{table}[ht]
 \caption{Computation time with respect to history size.
          Notation:
          $d$ is the number of timesteps used in the computation of the half derivative,
					$n$ is the number of particles,
					$t_{frame}$ is the average computation time per frame  without history terms, and
					$t_{history}$ is the average time per frame to compute the history terms.
         }
 \label{tab:perf}
 \begin{center}
   \begin{tabular}{c|r|r|r|r}
Experiment & $d$ & $n$ & $t_{frame}$ & $t_{history}$ \\ 
\hline
Bowling Ball&0   &20K &1.226 s & 0 ms \\
Bowling Ball&50  &20K &1.335 s & 4 ms \\
Bowling Ball&150 &20K &1.494 s &12 ms \\
Bowling Ball&500 &20K &1.627 s &40 ms \\
Vibration   &0   &20K &1.192 s & 0 ms \\
Vibration   &50  &20K &1.245 s & 4 ms \\
Vibration   &150 &20K &1.388 s &12 ms \\
Vibration   &500 &20K &1.539 s &40 ms \\
   \end{tabular}
 \end{center}
\end{table}

While only a given number of past timesteps are used in the computation of the fractional derivative term, the obtained value is an approximation of the true fractional derivative value and as such it can be as close as needed to the real value. The computation results in a weighted combination of past velocities that can be seen to resemble a local smoothing method; however, the fractional derivative formulation (Equation 3) enables the most interesting behavior of shear-thickening fluids  to emerge in simulation (Figures~\ref{fig:firstpic} and 
\ref{fig:monsters-compare}). These effects have not been demonstrated in simulation before.


As for limitations, although our model is able to imitate shear thickening fluids to a great extent, we have observed that in the vibration experiment the finger-like formations did not grow as tall as in the real experiment. This prevented the replication of finger-like formations falling and merging with each other forming bridges. 

Although we have not addressed shear thinning fluids and yield stress 
fluids in the scope of this paper, there is a wide literature on using fractional calculus for the simulation of these kinds of fluids as well. Yield stress fluids typically have hysteresis in the yield stress, although the hysteresis is in the opposite direction than for shear thickening fluids (higher viscosity after a history of resting). We believe these fluids can be simulated with variations of our model.
Another direction for future work would be to allow artists to manually edit history curves associated to the fluid, in order to allow the exploration of different effects for history-based fluids.

\section{Conclusions}

We have introduced a first attempt to visually simulate the dynamic behavior of shear thickening fluids.  
The proposed model is based on an efficient history-based stiffness term, which has showed to be essential for achieving discontinuous shear thickening behavior. 

The obtained results  successfully reproduce several behaviors that can be observed in real experiments.
Our results show that complex relationships between hysteresis, viscoelasticity and flows can be successfully reproduced, and the proposed techniques show great potential to be extended for addressing other types of mixtures.

The presented approach is relatively simple and efficient, and demonstrates a new technique for simulating unconventional behavior in a visually plausible way.



\bibliographystyle{IEEEtran}
\bibliography{15-arxiv-stf}
%

\end{document}